\documentclass{article}

\usepackage[a4paper, total={6in, 8in}]{geometry}

\usepackage[utf8]{inputenc}
\usepackage[utf8]{inputenc}
\usepackage{amsmath}
\usepackage{amssymb}
\usepackage{amsthm}
\usepackage{tikz}
\usepackage{graphicx}
\usepackage{mathtools}
\usepackage{multirow}
\usepackage{array}
\usepackage{verbatim}
\usepackage{adjustbox}
\usepackage[thinlines]{easytable}
\usepackage{hyperref}

\usepackage[bitstream-charter]{mathdesign}

\DeclareSymbolFont{usualmathcal}{OMS}{cmsy}{m}{n}
\DeclareSymbolFontAlphabet{\mathcal}{usualmathcal}

\begin{document}

\begin{center}{\Large \textbf{
Black Hole Mergers in Holographic Space-time Models of Inflation\\
}}\end{center}

\begin{center}
Anish Suresh\textsuperscript{1} and
Thomas Banks\textsuperscript{2$\star$}
\end{center}

\begin{center}
{\bf 1} Rutgers University, New Brunswick
\\
{\bf 2} Rutgers University, New Brunswick, New High Energy Theory Center
\\
${}^\star$ {\small \sf tibanks@ucsc.edu}
\end{center}

\begin{center}
\today
\end{center}

\begin{abstract}
    Holographic space-time, a theory of quantum gravity that generalizes string theory and quantum field theory, predicts black holes in the early matter-dominated era of its models of inflation. Before these black holes can decay, there is a chance that enough of these particles merge to produce radiation visible today in the Cosmic Microwave background. To discover if this is the case, we perform a rudimentary computer simulation. We show that no problematic black holes are formed by mergers in the Holographic Space-time models of inflation. However, we conclude that tiny bound structures containing black holes remnants form in this theory unconditionally. Since black hole decay products are mostly massive standard model particles, and perhaps their superpartners, the fate of these structures is a complicated dynamical problem that requires further study. It suggests the possibility of primordial structures on the order of the horizon size at the beginning of the radiation dominated era.  This is about $10^9\ L_P$ in the current model. 
\end{abstract}

\section{Introduction}
\label{sec:intro}

Ever since the conception of general relativity and quantum mechanics, physicists have been searching for a unified field theory: one theory that describes the three fundamental forces along with gravity. The standard model of particle physics successfully explains the electromagnetic, weak, and strong forces. However, the addition of gravity - namely, general relativity - has proven to be a challenge. Naively quantizing space-time gives rise to a myriad of mathematical problems, such as probabilities being larger than unity. Thus, a vastly different approach is needed to find a unified field theory. 

The biggest contender for such a theory is string theory, in which particles are replaced by one-dimensional objects such as strings. This theory has a mathematical framework that successfully incorporates gravity with the three fundamental forces. However, a flaw of string theory is that it does not describe our specific universe. There are two fundamental discrepancies between string theory and the real world: super-symmetry (SUSY) and cosmology. String theory finds an association between macroscopic space-time radius of curvature and almost exact supersymmetry, but the universe does not have this property. In addition, string models always have a non-positive cosmological constant, but the universe  appears to have a positive one. While string theory is mathematically consistent, it does not perfectly describe the real world. 

Holographic Space-time was introduced as a a more general framework for quantum gravity, which can accommodate a positive cosmological constant. Holographic space-time expands on string theory and quantum field theory (quantum field theory) using the strong holographic principle. In this theory, classical causal diamonds are replaced by quantum causal diamonds. That is, all measurements possible in a causal diamond in a $4$ dimensional space-time are encoded in a null foliation of the diamond boundary, with maximal area leaf: the holographic screen $A_{\diamond}$.  The von Neumann entropy of the quantum density matrix of an empty diamond is given by the $\frac{A_{\diamond}}{4G_N} = \frac{A_{\diamond}}{4}$, as we will use natural units ($\hbar = c = G_N = 1$) throughout this paper. Holographic space-time assumes this fact since it can be shown to produce Einstein's field equations from the local laws of thermodynamics, the purely geometrical Raychaudhuri equation, and Unruh's observation linking acceleration and temperature in weakly curved space-times\cite{jacobson1995thermodynamics}. 

HST actually makes the slightly stronger assumption that the Hilbert space of the diamond is finite dimensional. This was originally motivated by the assumption that the density matrix in a general diamond had to be maximally uncertain, but a more refined ansatz\cite{carlip1999black,solodukhin1999conformal,banks2021conformal} has recently been proposed.  These hypotheses map the causal structure and conformal factor of space-time into quantum concepts. 

In order to incorporate causality into the dynamics of HST, one introduces independent Hamiltonians along each timelike geodesic in a background classical space-time.  Following the logic of\cite{jacobson1995thermodynamics}, the classical background is thought of as describing the hydrodynamics of the quantum system one is trying to construct.  One breaks each geodesic up into nested intervals of proper time.  In the cosmological context, the past tip of each interval lies at the beginning of the universe, and the future tip of each successive interval is one Planck time in the future of that of the previous diamond.  Inside of a given diamond, time evolution of "generic states" is the analog of "one sided modular flow" in quantum field theory.  However, since all of the diamond Hilbert spaces are finite dimensional, this flow is just a time dependent unitary embedding of the Hilbert space of the smaller diamond into that of the next larger one.  A specific conjecture for the form of this embedding was recently given in\cite{banks2023hilbert}. We will only need to understand two things about this conjecture.  It postulated a universal form for the density matrix of a causal diamond $\rho_{\diamond} = e^{- K_{\diamond}} $ with
\begin{equation}
    \langle K_{\diamond} \rangle = \langle (K_{\diamond} - \langle K_{\diamond} \rangle)^2 \rangle = \frac{A_{\diamond}}{4}
\end{equation}
Secondly, localized excitations of energy $E$ as measured along the geodesic in the diamond, live in a constrained subspace of the Hilbert space, with of order $E\sqrt{A_{\diamond}}$ frozen q-bits.  The geodesic proper time scale for interactions that unfreeze these q-bits, and for all processes involving q-bits on the horizon is $\sqrt{A_{\diamond}}$.  This can be understood geometrically as the analog of the Milne redshift for near horizon processes in Minkowski space.  

To completely specify time evolution along a geodesic one must also evolve the system outside the current causal diamond at cosmological time $t$.  The fundamental conjecture of the HST formalism is that this is determined by the {\it Quantum Principle of Relativity} (QPR).  If we take two diamonds along two different geodesics they will have an overlap (which might be empty).  The QPR states that the maximal area causal diamond in the overlap must be identified with a tensor factor in each of the individual geodesic Hilbert spaces and that, for all initial conditions, the density matrices in these two tensor factors must have the same spectra.   

There is one set of initial conditions for which it is easy to satisfy the QPR.  We simply say that the state in each geodesic Hilbert space at any time is a generic one chosen from the universal density matrix.  Since these generic states live on the apparent horizon, their average energy at time $t$ scales like $E t^{-1} \langle K_{\diamond} \rangle = t^{-1} S \sim t$, where $S$ is the number of q-bits. Since there is no curvature scale in these relations we conclude that the resulting universe is automatically spatially flat.  The energy and entropy densities therefore scale like 
\begin{equation}
    \rho \sim t^{-2}, \ \ \ \ \sigma \sim t^{-2}.
\end{equation}
which gives an equation of state $p = \rho$.   We are free to stop the growth of the Hilbert space at any time, and let the system continue to evolve with a random Hamiltonian $K$ having the same two first moments.   This model describes a cosmology with scale factor $a(t) = \sinh^{1/3} (3t/R_I) $, where $\pi (R_I / L_P)^2$ is the entropy of the asymptotic density matrix.  In this model the QPR can be satisfied by saying that the density matrix on the overlap diamond is just the universal density matrix for diamonds of that area.  The model has no localized excitations so it really only has coarse grained observables.  

To construct a more realistic model, we imagine constrained initial conditions, such that the Hilbert space of the causal diamond slowly expands, but the system inside it just contains multiple copies of the above model with the same $R_I$ that are approximately non-interacting. These are, in conventional language, {\it inflationary horizon volumes}. In the post- inflation stages of this model \cite{banks2015cp, banks2017holographic, banks2018holographic, banks2020primordial, banks2021entropy, banks2022discretely}, individual inflationary horizon volumes are seen as black holes in the backward Milne coordinates of a slow roll geometry, as the horizon expands after inflation.  Most of these black holes decay, igniting the Hot Big Bang at a temperature around $10^{8} - 10^{10}$ GeV, but a fraction $\sim 10^{-8}$ of them carry the minimal charge under a discrete $Z_N$ gauge symmetry, and survive to be Primordial Black Hole (PBH) Dark matter, with mass of order $M_P = 1$. Prior to the Hot Big Bang, there is an era of early matter domination and primordial density fluctuations grow to be $o(1)$ before the black holes decay. One might worry, that the neutral black holes, whose initial mass is $\sim 10^6$ could combine to form black holes that would decay during an era where they would have left an imprint on observational data.  There are very strong constraints\cite{carr2019primordial,bernard2021constraints,carr2020primordial} on black hole number densities in the mass range whose Hawking lifetime is a few orders of magnitude below the current age of the universe. These problematic black holes could rule our model out.

In this paper, we study this problem numerically to determine whether that early growth of structure could lead to signals that could falsify the model. We simulate a toroidal lattice of black holes, incorporating periodic boundary conditions and an expanding universe to produce a homogeneous and isotropic cosmological space-time. Since black hole mergers are highly relativistic and one merger alone is a complicated process, we simplify our merger process to one of conservation of momentum. Through this process, we find that no mergers occur, meaning that Holographic space-time is not inconsistent with the real world. We then find that entire groups of black holes `stick' together, resulting in bound structures containing black hole remnants.

This paper is outlined as follows. Section \ref{sec:Code Structure and Parameters} details the assumptions and parameters that our code uses. Section \ref{sec:Results} explains the results of our simulations. Then, section \ref{sec:Macroscopic Behavior} contains analysis of the behavior of black holes on a macroscopic scale. Finally, we conclude our thoughts and provide commentary for the next steps of this project in section \ref{Conclusion}.

\section{Code Structure and Parameters}
    \label{sec:Code Structure and Parameters}

    To determine if most black holes decay before they merge, we wrote a program similar to an $N$-Body simulation on Jupyter Notebooks. The program starts with $N$ initial black holes with random initial velocities and masses, all evenly spaced in a lattice. Every time-step, we check to see if there exist pairs of black holes that are close enough to merge, and adjust their parameters accordingly. This is accompanied by keeping track of the expansion of the universe. In addition, the simulation topology is chosen as a 3-D torus to find a result for the whole universe limit computational resources. Below, we will formalize these notions. 

    \subsection{Parameters of the Simulation}
    \label{subsec:Parameters}

    \subsubsection{Velocity}

     We randomly generate velocities from a normal distribution with mean $0$ and standard deviation $\sigma_v = \frac{1}{10}$ (since we are using natural units, this is equivalent to $\frac{c}{10}$). This type of distribution is used, since we do not have any reason to believe the velocities are Poisson or any other distribution. In addition, the mean velocity was chosen to be $0$, as the black holes are free to move in both the negative and positive directions. Furthermore, $\sigma_v$ was chosen to be low enough that black holes with high velocities, which would require relativistic corrections and is impractical for large number of particles, are not simulated. In Python, we used the version 1.20.1 \texttt{numpy} package and the function \texttt{numpy.random.normal} to produce this distribution.

     \subsubsection{Mass}

    The mass distribution is similar: we randomly produce values from a normal distribution, now with mean $\bar{m} = 10^{6}$ \cite{carr2019primordial} and standard deviation of order $1$. The latter value comes is what the density fluctuations grow to be non-linear - the time of the universe we attempt to study \cite{banks2017holographic}. The mass distribution of the black holes is also calculated with the aforementioned \texttt{numpy.random.normal} function.
    

    \subsubsection{Initial Time}

    The standard equations of cosmological perturbation theory say that
    \begin{equation}
        \label{eq: mass fluc}
        \frac{\delta m}{m} =\epsilon \left(\frac{\delta H}{H}\right) t^{2/3},
    \end{equation}
    where $\epsilon \equiv \frac{-\dot{H}}{H^2}$. In our model $\frac{\delta H}{H} \sim m^{-1}$, the entropy/mass fluctuation of a single black hole.  Note that this does not have a factor of $\sqrt{\epsilon}$ familiar from field theory models of inflation.  In those models, that factor comes from the normalization of the fluctuating gravitational field, treated as a perturbative quantum field, whereas in HST models, the physical origin of fluctuations comes from the quantum fluctuations of Inflationary Black Holes. $\epsilon$ is not constrained by the ``Swampland'' ideas, which assume that CMB fluctuations are those of a quantum field.  Rather it is constrained by the requirement that the horizon expand rapidly enough during the slow roll era that the individual inflationary horizon volumes behave as isolated quantum systems.  This gives the constraint
    \begin{equation}
        \label{constraint}
        \epsilon > B ({\rm ln}\ (H^{-1}))^{-1}.
    \end{equation}
    The parameter $B$ here has to do with fast scrambling. If the natural time scale defined by a Hamiltonian is $T$ and the system is a fast scrambler, then quantum information introduced into one q-bit is distributed over all of the q-bits in a time of order $BT\ln{(S)}$, where $S$ is the number of q-bits and $B$ is a constant that varies from system to system. It's typical to assume that $B \sim 1$. Equation (\ref{constraint}) assumes that the horizon is a fast scrambler of quantum information. Given the observational constraint $\epsilon m \sim \epsilon H_I^{-1} \approx \epsilon H^{-1} = 10^{-5}$, where $H_I$ is the scale of the horizon during inflation, we find that $\epsilon \sim 0.1$. This is compatible with CMB data because as noted above, the ratio between scalar and tensor two point functions in HST models will scale like $\epsilon^2$ instead of $\epsilon$. 

    The initial time of the simulation $t_{0}$ can be found with \eqref{eq: mass fluc} with $\delta m/m \sim 1$, as mentioned prior. Plugging in these values produces $t_0 = 10^{\frac{21}{2}}$.

    \subsubsection{Spatial Distribution and Density}

    A lattice is specifically chosen due to its isotropy and homogeneity on large scales. These assumptions are chosen to easily satisfy the consistency conditions of HST, which state that if two trajectories intersect (and therefore share information), their corresponding density matrices must have the same spectra in this region \cite{banks2017holographic}. It has been postulated that homogeneity is a prerequisite for this complex condition \cite{banks2017holographic}. The addition of isotropy is guaranteed to satisfy the coarse-grained consistency condition, making the lattice a helpful setting to study. Furthermore, we limit the size of the lattice to a `cube' or \textit{cell}, and make the metric a 3D torus. This process accounts for a large lattice structure, representing the universe, while keeping the computing power minimal. More information of this metric can be found in section \ref{subsec:Periodic Boundary Conditions}.

    We take the initial black hole number density in the post slow roll era to be just small enough that the black holes do not immediately recombine into a single horizon filling black hole. We estimate that this density is $ n = Cm^{-3} $ with $C \in [10^{-3},10^{-1}]$, after which the density falls as $n = Cm^{-3}t^{-2}$.  The reason for this initial condition is that, according to the logic of HST models, any universe with localized excitations comes from an improbable initial condition.  We want to find the least improbable set of initial conditions that could create a universe like the one we see. 
        
    Since $n$ can also be written as $NL^{-3}$, where $N$ is the number of black holes in the cell and $L$ is the length of one side of the cell. We can derive the starting length $l$ between adjacent black holes at $t_{0}$, giving us

    \begin{equation}
        \label{eq:latticespacing}
        l = \sqrt[3]{\frac{\bar{m}^{3}t_{0}^{2}}{C}},
    \end{equation}
    from the relation $L = \sqrt[3]{N}l$. 

    \subsubsection{Updating Position and Velocity}

    We can use the numerical integration technique Leapfrog Integration to update the positions and velocities of the particles every time-step. This technique, which can be utilized to solve any differential equation of the form $\ddot{x} = f(x)$, has the following steps (for each time-step):

    \begin{equation}
        \begin{split}
            \Delta x = v\Delta t + \frac{1}{2}a_{\text{old}}(\Delta t)^{2}, \hspace{1cm} \Delta v = \frac{1}{2}a_{\text{old}}\Delta t, \hspace{1cm} \Delta v = \frac{1}{2}a_{\text{new}}\Delta t,
        \end{split}
    \end{equation}

    where $a_{\text{new}}$ is the newly updated acceleration based on the change in position. More information about this technique can be found at \cite{birdsall2004plasma}. We are able to use this method because the dynamics of each black hole is solely governed by the (classical) gravitational force it feels.

    \subsection{Periodic Boundary Conditions}
    \label{subsec:Periodic Boundary Conditions}

    Since our topology is $\mathbb{T}^{3} \equiv S^{1} \times S^{1} \times S^{1}$, we have to account for some of its properties. The most obvious consequence of this topology is that if a particle leaves the cell, then the particle instantaneously comes back in through the side diametrically opposite. Such a condition helps ensure homogeneity. 
    
    Another such property is how particles feel a force from a \textit{boundary} black hole (i.e. black holes that exist on the boundary of our toroidal lattice) from two directions. For example, if there exists a black hole in the middle of the cell, and there exists a black hole on the edge directly to the left of the former particle, then the black hole in the middle feels a force directly from the left and right.

    \usetikzlibrary{arrows, arrows.meta}

    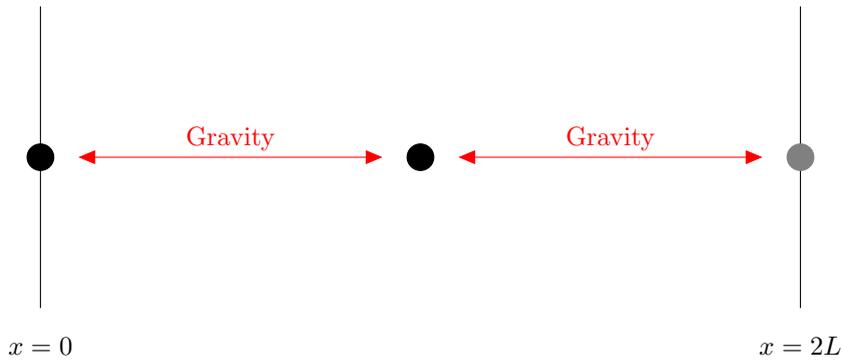
\begin{figure}[h!]
    \begin{center}
    \begin{tikzpicture}
     
    \draw[] (-5,-2) -- (-5,2);
    \draw[] (5, -2) -- (5, 2);
    \filldraw[black] (0,0) circle (5pt);

    \filldraw[black] (-5,0) circle (5pt);
    \filldraw[gray] (5,0) circle (5pt);

    \draw[>=triangle 45, <->, red] (-4.5,0) -- (-0.5,0);

    \draw[>=triangle 45, <->, red] (4.5,0) -- (0.5,0);

    \draw[red] (-2.5, 0.25) node {Gravity};
    \draw[red] (2.5, 0.25) node {Gravity};

    \draw[] (-5, -2.5) node {$x = 0$};
    \draw[] (5, -2.5) node {$x = 2L$};
    
    \end{tikzpicture}
    \caption{An example to illustrate how forces travel in $\mathbb{T}^{1}$. There is one boundary black hole on the left and one interior black hole exactly in the middle. Due to the topology, the former can be thought of as being on the right, which we have modelled with a mirror black hole on the right. Forces have periodic boundary conditions as well, so this topology keeps the interior black hole in place, which in turn keeps the stationary black hole in place. This scenario is drastically different than one in $\mathbb{R}^{1}$, in which the black holes would have gravitated towards each other. Identically, we can think of the force on the right to be from the same black hole, but in the opposite boundary side. }
    \label{fig:PBCs} 
    
    \end{center}
    \end{figure}

    To simplify our process a bit, we utilize the method of images and create \textit{mirror} black holes that are placed diametrically opposite to boundary black holes (it is important to note that they are two different representations of the same black hole). More specifically, each boundary black hole will have exactly one mirror black hole, with the same mass, velocity, and acceleration in our simulation\footnote{This is actually not true: each boundary black hole on an edge of our cube has 3 mirrors and a boundary black hole on a vertex has 7 mirrors (namely, the other vertices). However, this level of detail makes our program much more complicated with little benefit, so we will work with our approximation.}. An important note about acceleration is that a boundary black hole will feel gravity from particles near it and from particles near its mirror (since its mirror is a representation of itself). This detail applied to our earlier scenario can be seen in Figure \ref{fig:PBCs}.
    
\begin{figure}[h!]
    \begin{center}
    \begin{tikzpicture}
     
    \draw[] (0,0) -- (0,4);
    \draw[] (0, 0) -- (4, 0);
    \draw[] (0, 0) -- (1.414, 1.414);
    \draw[] (4,0) -- (4,4);
    \draw[] (0,4) -- (4,4);
    \draw[] (1.414, 1.414) -- (1.414 + 4, 1.414);
    \draw[] (4, 0) -- (4 + 1.414, 1.414);
    \draw[] (0, 4) -- (1.414, 4 + 1.414);
    \draw[] (1.414, 4 + 1.414) -- (1.414 + 4, 4 + 1.414);
    \draw[] (1.414, 1.414) -- (1.414, 4 + 1.414);
    \draw[] (4 + 1.414, 1.414) -- (4 + 1.414, 4 + 1.414);
    \draw[] (4, 4) -- (1.414 + 4, 4 + 1.414);
    
    \filldraw[black] (0,0) circle (5pt);
    \filldraw[black] (2,0) circle (5pt);
    \filldraw[black] (4,0) circle (5pt);

    \filldraw[black] (0,0) circle (5pt);
    \filldraw[black] (0,2) circle (5pt);
    \filldraw[black] (0,4) circle (5pt);

    \filldraw[black] (0,0) circle (5pt);
    \filldraw[black] (1.414/2,1.414/2) circle (5pt);
    \filldraw[black] (2.828/2,2.828/2) circle (5pt);

    \filldraw[black] (4,2) circle (5pt);
    \filldraw[black] (2,2) circle (5pt);
    \filldraw[gray] (1.414,1.414 + 4) circle (5pt);

    \filldraw[black] (1.414/2,1.414/2 + 2) circle (5pt);
    \filldraw[black] (1.414/2,1.414/2 + 4) circle (5pt);
    \filldraw[black] (2,4) circle (5pt);
     \filldraw[black] (1.414/2 + 2,1.414/2) circle (5pt);

    \filldraw[gray] (1.414,1.414 + 2) circle (5pt);
    \filldraw[gray] (1.414 + 2,1.414) circle (5pt);
    \filldraw[gray] (1.414 + 4,1.414) circle (5pt);

    \filldraw[gray] (1.414 + 4,1.414 + 4) circle (5pt);
    \filldraw[gray] (1.414 + 2,1.414  + 4) circle (5pt);
    \filldraw[gray] (1.414 + 4,1.414 + 2) circle (5pt);

    \filldraw[gray] (4,4) circle (5pt);
    \filldraw[gray] (1.414 + 2,1.414 + 2) circle (5pt);
    \filldraw[gray] (1.414/2 + 4,1.414/2 + 2) circle (5pt);
    \filldraw[gray] (1.414/2 + 4,1.414/2 + 4) circle (5pt);

    \filldraw[gray] (1.414/2 + 4,1.414/2) circle (5pt);
    \filldraw[gray] (1.414/2 + 2,1.414/2 + 4) circle (5pt);

    \filldraw[blue] (1.414/2 + 2,1.414/2 + 2) circle (5pt);

    \draw[] (-0.5,-0.5) node {$(0,0)$};
    \draw[] (1.414 + 4 + 0.5,1.414 + 4 + 0.5) node {$(2l,2l)$};
    
    \end{tikzpicture}
    \caption{The 3D toroidal lattice structure at $t = t_{0}$ with $N = 27$. Here, we have $13$ boundary black holes (in black), which we model on the corresponding diametrically opposite side as $13$ mirror black holes (in gray). We also have one interior black hole (in blue).}
    \label{fig:N=3 lattice} 
    
    \end{center}
    \end{figure}
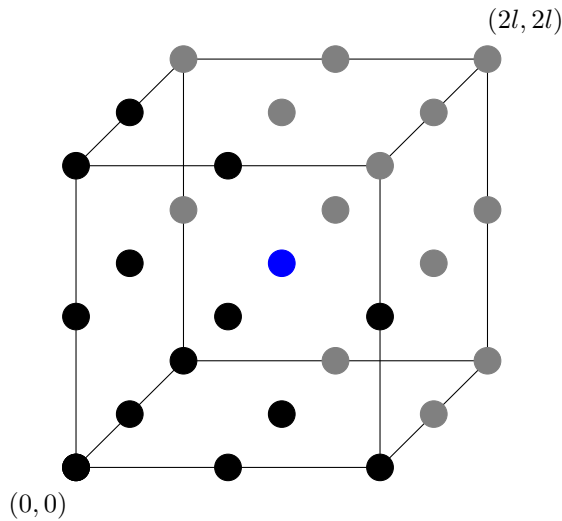

    Out of the $N$ points in our simulation, there exist $\frac{1}{2}[N - (\sqrt[3]{N} - 2)^{3}]$ mirrors. Thus, we are simulating $\frac{1}{2}[N + (\sqrt[3]{N} - 2)^{3}]$ unique black holes. Figure \ref{fig:N=3 lattice} shows what our toroidal lattice looks like at the start when $N = 27$\footnote{The perceptive reader would have noticed the slight issue with Figure \ref{fig:N=3 lattice}. Earlier, we assumed that $L = \sqrt[3]{N}l$, which is not $2l$ when $N = 27$. Since we are working with a large $N$, we do not have to worry about this discrepancy, since the factor $\sqrt[3]{N}/(\sqrt[3]{N} - 1)$ that should have been in \eqref{eq:latticespacing} is approximately $1$ when $N$ is high.}.

    \subsection{Merger Process}
    \label{subsec:Merger Process}

    It is well known that the merging of black holes is a complicated and highly relativistic process. With high accuracy, a simulation of even one pair coalescing together is extremely detailed. Since we are working with a lot of black holes, we will be simplifying this process to one of momentum conservation\footnote{We will not be working with Kerr black holes or binary systems, so we shall only consider conservation of linear momentum.}. Namely, we shall think of two black holes merging as a perfectly inelastic collision, resulting in the following newly formed particle's parameters 

    \begin{equation}
        \label{momentumconserve}
        m = m_{i} + m_{j}, \hspace{1cm} x = \frac{1}{2}(x_{i} + x_{j}), \hspace{1cm} v = \frac{m_{i}v_{i} + m_{j}v_{j}}{m_{i} + m_{j}},
    \end{equation}

    where the first and last equations come from conservation of momentum. The position is deemed to be the midpoint of the two black holes before the merge. 
    
    The condition to merge is when two black holes get closer than $10m$, the Schwarzschild radius, of each other, where $m$ is the mass of either black hole. That is, black holes $i$ and $j$ merge when:

    \begin{equation}
        \label{conditionofmerge}
        \text{distance}(i,j) \leq \text{min}(10m_{i}, 10m_{j})
    \end{equation}

    In addition, if $i$ and $j$ are both boundary and/or mirror black holes, then their mirrored particles, denoted as $i^{*}$ and $j^{*}$, will also merge. Obviously, this is a rudimentary approximation to the actual black hole merger process. We believe that, given our conclusions, it is adequate.  We will find that the black holes never get close enough to merge, so all the complications of the merger process never come into play.

    Before we analyze results, we must consider one other core component of this simulation.

    \subsection{Expansion of the Universe}
    \label{subsec:Expansion of the Universe}

    Since we are simulating the matter dominated era of the universe, the normalized Friedmann's equation tells us that

    \begin{equation}
        \label{eq:friedmanneq}
        \frac{H^{2}}{H_{0}^{2}} = \Omega_{r}a^{-4} + \Omega_{m}a^{-3} + \Omega_{k}a^{-2} + \Omega_{\Lambda} \approx \Omega_{m}a^{-3},
    \end{equation}
    
    where we rightfully assumed that $\Omega_{r}, \Omega_{\Lambda} \approx 0$ due to the era of the universe, and it has been shown that $\Omega_{k} \approx 0$ in the early universe \cite{baumann2022cosmology}. Through integration, we arrive at $a(t) \propto t^{\frac{2}{3}}.$ As a result, our expansion is governed by $a = \left(\frac{t}{t_{0}}\right)^{2/3}$. In the first time-step in which $a \geq 2$, we make $a = 1$ again and double the size of our lattice spacing (which in turn doubles the physical distance between all particles). We allow this process to continue $9$ more times; that is, we allow the physical volume of our cube to grow up to $2^{30}$ times the start volume after considerable time has passed. We stop at this point, as the probability of merging after $10$ expansions is minuscule. 

    \subsection{Time Step}
    \label{subsec:Time Step}
    Finding an appropriate value $\Delta t$ for the time-step depends on multiple factors, such as time of the simulation and expansion of the universe. Theoretically, the smaller $\Delta t$ is, the more accurate the simulation is. However, this may pose a problem if the `ending' time of the simulation is large. Such a time in this scenario represents the time it takes for black holes to decay. We know that this decay time is approximately

   \begin{equation}
    \label{eq:decaytime}
       t_{d} = 2^{10}\pi g^{-1}10^{19} \approx 2^{10} \pi 10^{16},
   \end{equation}

   where $g$ represents the number of particle species and is approximately equal to $10^{3}$. This means that if $\Delta t$ is as `small' as $10^{3}$, then the simulation will take ~$\frac{10^{19} - 10^{21/2}}{10^{3}} \sim 10^{16}$ time steps, which is an extremely large number. This value can be drastically reduced by safely assuming that most of mergers (if any) will happen before the universe expands even once (due to the increase in distance and decrease in acceleration). Thus, a good approximation for the `end' of our simulation is the time it takes to expand. To find this time, we use the previously derived equation for the scale factor, finding $t \sim 6 \times 10^{10}$. Thus, the number of time steps needed is $\frac{10^{10}}{\Delta t}$. We can safely set $\Delta t$ to some value between $10^{5} - 10^{7}$; this is large enough that results are obtainable and small enough that it doesn't negatively impact the simulation\footnote{If two particles are right outside of the merging threshold, then they can move straight past each other without merging when $\Delta t$ is large. This happens due to the high acceleration they obtain from being near each other, which causes a large change in position.}.

\section{Results Inside Each Cell}

    \label{sec:Results}

    After running the simulation, the code for which is available on \href{https://github.com/EulersLoveChild/black-hole-mergers}{GitHub}, there seem to be $0$ mergers occurring. To make sure our results were not a result of the input parameters chosen, we utilized many possible values for $C$, the standard deviation of the velocity $\sigma_v$, and $\Delta t$. Table \ref{tbl 1} provides a list of values used. This strongly suggests that no merger will occur, a result independent of the initial conditions. It is worth noting that we limited $N$ to $216$ due to the complexity of the program. Furthermore, we set $C = 10^{19}, 10^{20},$ and $10^{21}$ (which significantly decreases the lattice spacing according to \eqref{eq:latticespacing}) to make sure the code was running correctly. Here, we noticed that black holes merged together in this setting.

    \renewcommand{\arraystretch}{1.5}

    \begin{table}[h!]
    \centering
    \begin{tabular}{|c|c|}
    \hline
    Parameter  & Values  \\ \hline
    $C$        & $10^{-3},5\times10^{-3},10^{-2},5\times10^{-3},10^{-1}$ \\ \hline
    $(\sigma_v)^{-1}$ & $20, 16, 10, 8, 4$ \\ \hline
    $\Delta t$ & $10^{5},5\times10^{5},10^{6},5\times10^{6},10^{7}$\\ \hline
    \end{tabular}
    \caption{A table of the values we considered for each parameter. No combination of these values resulted in any merge.}
    \label{tbl 1}
    \end{table}

    Let us consider a hypothetical scenario to assess the legitimacy of our results. Suppose there are two black hole of mass $10^{7}$ separated by the distance $l$. Then, note that they merge when the distance between them is less than $10^{8}$. In addition, suppose that both start out with speed $\frac{1}{4}$ and with velocities such that they are heading towards each other. Finally, since the maximum acceleration they will feel is $\frac{10^{7}}{10^{16}} = 10^{-9}$, we assume they have a constant acceleration of $10^{-9}$ (again in a way that they are moving towards each other, with increasing speeds). The reason why we are analyzing this scenario is to determine the time taken for two black holes to merge, when they start in the toroidal lattice (given our artificially inflated initial parameters such as velocity, acceleration, mass, etc.). With $C = 0.1$, we get $l \sim 2 \times 10^{13}$, so $t \sim 2 \times 10^{11}.$ This value is almost double the time it takes to expand the universe, even with unrealistically high parameters. That is, the universe expands before the black holes merge, meaning that the black holes are further apart. These calculations suggest that probability of black holes merging before the universe expands is low.

   \section{Macroscopic Behavior}
   \label{sec:Macroscopic Behavior}

   Up until now, we have been analyzing the behavior of black holes in a tiny portion of the Horizon volume and arrived at the conclusion that black holes do not coalesce in this microscopic setting. However, we are yet to consider the behavior of each cell of black holes; that is, such groups may become bound together.

   This can be determined by comparing two times, vaguely similar to our earlier work. Like before, we will need to calculate the time for the Horizon radius to double. The second time we need is new: the time it takes for mass on the outskirts of the Horizon volume to reach the center via gravitational collapse. For simplicity, we assume that the Horizon volume is spherical. This is then equivalent to a collapse of a Newtonian shell.

   \subsection{Expansion Time}
   \label{subsec:Expansion Time}

   To find the first time, we need an expression for the radius. We know that the coordinate of the horizon $K$ grows with the following differential equation:

   \begin{equation}
       \label{eq:kdot}
       \dot{K} = \frac{1}{a}.
   \end{equation}

   Using Friedmann's Equation, we can solve for $K$ and in turn the physical distance $R_H$, which is simply the coordinate size times the expansion coefficient:

   \begin{equation}
   \begin{split}
       \label{eq:Kvsa}
       R_{H}(a) & = R_{0} + 2\sqrt{\frac{3}{8\pi\rho_{0}}}a^{3/2} = 10\bar{m} + 2\bar{m}\sqrt{\frac{3*10^{3}}{8\pi}}a^{3/2}.
   \end{split}
   \end{equation}
    
    In addition, we simplified the above equation with $\rho_{0} = \frac{\bar{m}}{(10\bar{m})^{3}} = 10^{-3}\bar{m}^{-2}$ and $R_{0} = 10\bar{m}$, which is the Schwarzschild radius of a black hole with average mass $\bar{m}$. Our analysis happens when $a \approx 10^{5}$, meaning that we can safely ignore the first term.
   
   It is easy to see that $R_{H}$ doubles when $a$ becomes $2^{2/3}10^{5}$. To determine how much time this takes, we can integrate Friedmann's equation. This straightforward calculation gives us $t_{\text{expansion}} = 2.303 \times 10^{14}.$

    \subsection{Collapse Time}
    \label{subsec:Collapse Time}

    Now, we need to find the time it would take for mass at the edge of the horizon volume to reach the center. The equation of motion of such a mass is
    
    \begin{equation}
    \label{eq:Newtonianeqforcollapse}
        \ddot{x} = -\frac{G(2M)}{x^{2}},
    \end{equation}
    
    where $2M = \frac{8}{3}\pi R^{3}_{H}\rho$ is the mass at the center. Luckily, this equation of motion is 1-dimensional and autonomous, so the time it takes for the particle to start at $x = R_{H}(a = 10^{5})$ and end at $x = 0$ can be solved for. This derivation is identical to the one for freefall time, the time it takes for a gas to collapse due to its own gravity. As explained in \cite{keeton2014principles}, this has the form $t_{\text{ff}} = \sqrt{\frac{3\pi}{32\rho}} = 5.427 \times 10^{14},$ when plugged in our values. Notice that $t_{\text{ff}} > t_{\text{expansion}}$\footnote{This equation is the general expression for free fall time with mass $M$, but since we have mass $2M$, the 32 should be replaced by a 64. Regardless, the inequality holds true.}. 

    Before we reach a conclusion, a more careful look at the expression for free fall time is necessary. This variable is inversely related to $M$ (so, also $\rho$), and we know the mass is not fixed in our scenario. If we recall, $M$ is dependent on $R_{H}$ and $\rho$, which are both functions of time. Thus, our equation of motion becomes

    \begin{equation}
        \label{eq:EOM for a Newtonian Shell}
        \ddot{x} = -\frac{8\pi G R^{3}_{H}(t)\rho(t)}{3x^{2}}.
    \end{equation}

    Now, we get a second-order non-autonomous differential equation, which is much harder to solve by hand, so we can do this numerically. This involves first updating $R_{H}$ and $\rho$, which will help us update the position $x$.

    For the former two, we will be using Runge-Kutta 4, a popular numerical integration technique for first-order differential equations. For example, if we wanted to update $x$ based on its derivative $f(x)$, then we would first create four parameters\footnote{These are the parameters if $\dot{x}$ is purely a function of $x$.}:
    
    \begin{equation}
        \label{eq:RK4 ks}
        k_{1} = f(x), \hspace{0.2cm} k_{2} = f\left(x + dt\left(\frac{k_{1}}{2}\right)\right), \hspace{0.2cm} k_{3} = f\left(x + dt\left(\frac{k_{2}}{2}\right)\right), \hspace{0.2cm} k_{4} = f(x + dt(k_{3}))
    \end{equation}

    Now, the updated position is
    
    \begin{equation}
        \label{eq:RK4}
        x_{\text{new}} = x_{\text{old}} + \frac{1}{6}dt(k_{1} + 2k_{2} + 2k_{3} + k_{4}).
    \end{equation}
    
    More information about Runge-Kutta 4 can be found here \cite{butcher2007runge}.

    To effectively use this method, notice that both $R_{H}$ and $\rho$ can be written as functions of $a$. With $dt = 10^{7}$ (which is reasonable, since $dt \ll t_{\text{ff}}$), we get that $t_{\text{collapse}} = 2.989 \times 10^{13}$, as seen in Figure \ref{fig:M,x vs t} (the code for this calculation can be found on \href{https://github.com/EulersLoveChild/black-hole-mergers}{GitHub}). With this more accurate model, we find that the time needed for a mass on the boundary of a Newtonian shell is considerably less than our expansion time, meaning that groups of black holes are bound together. It is important to understand that this result was not dependent on our choice of initial conditions; these bound structures are products of the theory of HST.

    \begin{figure}[h!]
    \centering
    \includegraphics[scale=0.5]{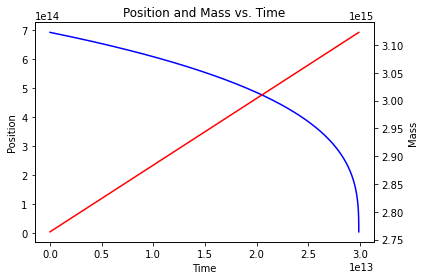}
    \caption{Position $x$ (blue) and Mass $M$ (red) with respect to time. Clearly, $x$ reaches 0 around $t = 3 \times 10^{13}$. Notice that mass increases with time. In fact, it is linear.}
    \label{fig:M,x vs t}
    \end{figure}
    
    It is worth noting that $t_{\text{collapse}}$ being smaller than $t_{\text{ff}}$ makes physical sense. The reason can be realized by observing $M$ with respect to time. We already know that

    \begin{equation}
    \label{eq:Mvsa}
        M(a) = \frac{8}{3}\pi R^{3}_{H}(a)\rho(a) \propto a^{3/2}
    \end{equation}

    Since $a$ increases, $M$ increases as well, strengthening the force of gravity. Thus, $t_{\text{collapse}}$ is rightfully smaller than $t_{\text{ff}}$. The full evolution of $M$ with respect to time can be seen in Figure \ref{fig:M,x vs t}. Notice the linearity of the red plot. This is in agreement with our work, as $a \propto t^{2/3}$.

\section{Conclusion}
\label{Conclusion}
The main conclusion of our paper is that the Holographic space-time model of inflation survives as a model of the early universe.  As outlined in previous work\cite{banks2015cp, banks2017holographic, banks2018holographic, banks2020primordial, banks2021entropy, banks2022discretely} it gives us an economical explanation of the CMB fluctuations\footnote{We do not yet have a precise calculation of the tensor to scalar ratio, but work is in progress on that.}, baryogenesis, and PBH dark matter.  The model is fully quantum mechanical, causal and unitary, and has no singularities or Trans-planckian problems.  It gives predictions for the detailed form of the tensor fluctuation spectrum and for non-Gaussian fluctuations, which differ from those of field theoretic models.  The current paper shows that no problematic black holes, whose decays could have given signals that falsified the model, are formed during the early matter dominated era.  Intriguingly we have also found that it is likely that bound structures do form during this era.  Most of the black holes making up these structures decay, but most of their decay products are actually massive standard model particles, and their superpartners, if those have masses less than $\sim 10^{13}$ GeV.  The size of these structures when they are formed is about $10^{28}$. This is roughly $10^{-4}$ cm, indicating their large size on microscales. The structures also contain a sprinkling of Planck mass stable PBHs,  Thus, it seems conceivable that these early structures could have an interesting effect on the evolution of structure in the early universe.  One would have to understand the rate at which the massive particles cool and whether they remain bound to the PBH clusters, perhaps forming the nuclei around which early galaxies coalesce.

\section*{Acknowledgements}

\paragraph{Funding information}
We thank Profs. A. Brooks and M. Buckley for useful conversations about galaxy formation simulations. This work was supported in part by the U.S. Dept. of Energy under Grant DE-SC0010008.

\bibliography{ref}

\providecommand{\href}[2]{#2}\begingroup\raggedright\begin{thebibliography}{10}

\bibitem{jacobson1995thermodynamics}
T.~Jacobson, ``Thermodynamics of spacetime: the einstein equation of state,'' {\em Physical Review Letters} {\bfseries 75} no.~7, (1995) 1260, \href{https://arxiv.org/abs/gr-qc/9504004}{{\ttfamily arXiv:gr-qc/9504004 [gr-qc]}}.

\bibitem{carlip1999black}
S.~Carlip, ``Black hole entropy from conformal field theory in any dimension,'' {\em Physical Review Letters} {\bfseries 82} no.~14, (1999) 2828, \href{https://arxiv.org/abs/hep-th/9812013}{{\ttfamily arXiv:hep-th/9812013 [hep-th]}}.

\bibitem{solodukhin1999conformal}
S.~N. Solodukhin, ``Conformal description of horizon's states,'' {\em Physics Letters B} {\bfseries 454} no.~3-4, (1999) 213--222, \href{https://arxiv.org/abs/hep-th/9812056}{{\ttfamily arXiv:hep-th/9812056 [hep-th]}}.

\bibitem{banks2021conformal}
T.~Banks and K.~M. Zurek, ``Conformal description of near-horizon vacuum states,'' {\em Physical Review D} {\bfseries 104} no.~12, (2021) 126026, \href{https://arxiv.org/abs/2108.04806}{{\ttfamily arXiv:2108.04806 [hep-th]}}.

\bibitem{banks2023hilbert}
T.~Banks, ``Hilbert bundles and holographic space-time models,'' \href{https://arxiv.org/abs/2306.07038}{{\ttfamily arXiv:2306.07038 [hep-th]}}.

\bibitem{banks2015cp}
T.~Banks and W.~Fischler, ``Cp violation and baryogenesis in the presence of black holes,'' \href{https://arxiv.org/abs/1505.00472}{{\ttfamily arXiv:1505.00472 [hep-th]}}.

\bibitem{banks2017holographic}
T.~Banks and W.~Fischler, {\em {Holographic Inflation Revised}}, \href{https://dx.doi.org/10.1017/9781316535783.013}{pp.~241--262}.
\newblock {Cambridge University Press Cambridge, UK}, 2017.
\newblock \href{https://arxiv.org/abs/1501.01686}{{\ttfamily arXiv:1501.01686 [hep-th]}}.

\bibitem{banks2018holographic}
T.~Banks and W.~Fischler, ``The holographic spacetime model of cosmology,'' \href{https://dx.doi.org/10.1142/S0218271818460057}{{\em International Journal of Modern Physics D} {\bfseries 27} no.~14, (2018) 1846005}, \href{https://arxiv.org/abs/1806.01749}{{\ttfamily arXiv:1806.01749 [hep-th]}}.

\bibitem{banks2020primordial}
T.~Banks and W.~Fischler, ``Primordial black holes as dark matter,'' \href{https://arxiv.org/abs/2008.00327}{{\ttfamily arXiv:2008.00327 [hep-th]}}.

\bibitem{banks2021entropy}
T.~Banks and W.~Fischler, ``Entropy and black holes in the very early universe,'' \href{https://arxiv.org/abs/2109.05571}{{\ttfamily arXiv:2109.05571 [hep-th]}}.

\bibitem{banks2022discretely}
T.~Banks and W.~Fischler, ``Discretely charged dark matter in inflation models based on holographic space-time,'' \href{https://dx.doi.org/10.3390/universe8110600}{{\em Universe} {\bfseries 8} no.~11, (2022) 600}, \href{https://arxiv.org/abs/2209.08361}{{\ttfamily arXiv:2209.08361 [hep-th]}}.

\bibitem{carr2019primordial}
B.~Carr and F.~K{\"u}hnel, ``Primordial black holes with multimodal mass spectra,'' \href{https://dx.doi.org/10.1103/PhysRevD.99.103535}{{\em Physical Review D} {\bfseries 99} no.~10, (2019) 103535}, \href{https://arxiv.org/abs/1811.06532}{{\ttfamily arXiv:1811.06532 [astro-ph.CO]}}.

\bibitem{bernard2021constraints}
B.~Carr, K.~Kohri, Y.~Sendouda, and J.~Yokoyama, ``Constraints on primordial black holes rept,'' \href{https://dx.doi.org/10.1088/1361-6633/ac1e31}{{\em Prog. Phys} {\bfseries 84} (2021) 116902}, \href{https://arxiv.org/abs/2002.12778}{{\ttfamily arXiv:2002.12778 [astro-ph.CO]}}.

\bibitem{carr2020primordial}
B.~Carr and F.~K{\"u}hnel, ``Primordial black holes as dark matter: recent developments,'' \href{https://dx.doi.org/10.1146/annurev-nucl-050520-125911}{{\em Annual Review of Nuclear and Particle Science} {\bfseries 70} (2020) 355--394}, \href{https://arxiv.org/abs/2006.02838}{{\ttfamily arXiv:2006.02838 [astro-ph.CO]}}.

\bibitem{birdsall2004plasma}
C.~K. Birdsall and A.~B. Langdon, \href{https://dx.doi.org/10.1201/9781315275048}{{\em Plasma physics via computer simulation}}.
\newblock CRC press, 2004.

\bibitem{baumann2022cosmology}
D.~Baumann, \href{https://dx.doi.org/10.1017/9781108937092}{{\em Cosmology}}.
\newblock Cambridge University Press, 2022.

\bibitem{keeton2014principles}
C.~Keeton, \href{https://dx.doi.org/10.1007/978-1-4614-9236-8}{{\em Principles of Astrophysics}}.
\newblock Springer, 2014.

\bibitem{butcher2007runge}
J.~Butcher, ``Runge-kutta methods,'' \href{https://dx.doi.org/10.4249/scholarpedia.3147}{{\em Scholarpedia} {\bfseries 2} no.~9, (2007) 3147}.

\end{thebibliography}\endgroup

\bibliographystyle{utphys.bst.txt}

\end{document}